\documentclass{cimento}
      [1999/12/01 v1.4c Il Nuovo Cimento]
\usepackage{bm}
\usepackage{graphicx}
\usepackage{latexsym}
\usepackage{amsmath}
\usepackage{amsfonts}
\usepackage{amssymb}
\newcommand{\beq}{\begin{equation}}
\newcommand{\eeq}{\end{equation}}
\newcommand{\beqy}{\begin{eqnarray}}
\newcommand{\eeqy}{\end{eqnarray}}

\newcommand{\as}{\alpha_s}

\title{Bound states of UED level-1 KK quarks  at the Linear Collider}
\author{O.~Panella\from{ins:x}
        \atque
N.~Fabiano\from{ins:x}
}
\instlist{\inst{ins:x} INFN Sezione di Perugia, Via A. Pascoli I-06123 Perugia, Italy}
\PACSes{\PACSit{12.60.-i}{Models Beyond the Standard Model}
\PACSit{11.10.St}{Bound states effects}}

\begin{document}
\maketitle

\begin{abstract}
We study the formation and detection at the next linear $e^+e^-$ collider of
bound states of level-1 quark Kaluza-Klein excitations ${\cal B}_{KK}$ within a
scenario of universal extra-dimensions (UED). 
\end{abstract}

\PACSes{12.60.-i, 11.10.St, 14.80.-j}

\maketitle
\section{Introduction}
\label{sec:intro}

It is well known that as early as 1921 Theodore Kaluza 
proposed a theory that was intended to unify gravity and
electromagnetism by considering a space-time with one
extra \emph{space-like} dimension~\cite{Kaluza:1921tu}. A
few years later Oscar Klein proposed that the extra space
dimension (the fifth dimension) is in reality compactified
around a circle of very small radius~\cite{Klein:1926tv}.
These revolutionary ideas have thereafter been ignored for
quite some time. However recent  developments in the field
of string theory have suggested again the possibility that
the number of space time dimensions is actually different
from $D=4$ (indeed string theory models require $D= 11 $,
i.e. seven additional dimensions). In 1990 it was
realized~\cite{Antoniadis:1990ew}  that string theory
motivates scenarios in which the size of the extra
dimensions could be  as large as $R\approx 10^{-17}$
cm (corresponding roughly to electro-weak  energy
scale ($\approx$ TeV) contrary to naive expectations  which
relate them to  a scale of the order of the Planck length
$L_P\approx 10^{-33}$ cm (corresponding to the
Planck mass $M_P=\sqrt{\hbar c/G} \approx 10^{19}$ GeV).
See also~\cite{Antoniadis:1998ig}.

Subsequently two approaches have been developed to discuss the observable
effects of these, as yet, hypothetical extra dimensions. One possibility is to
assume that the extra space-like dimensions are flat and compactified to a
``small'' radius. This is the so called ADD model~\cite{ArkaniHamed:1998rs}
where only the gravitational interaction is assumed to propagate in the 
extra-dimension. A second possibility is contemplated in the Randall-Sundrum
type of models where the extra dimensions do have curvature and are embedded in
a warped geometry~\cite{Randall:1999ee,Randall:1999vf}.

Universal extra-dimensional models were introduced in
ref.~\cite{Appelquist:2000nn} and are characterized, as opposed to the ADD
model, by the fact that all particles of the Standard Model (SM) are allowed to
propagate in the (flat) extra space dimensions, the so called \emph{bulk}. Here
to each SM particle $X^{(0)}$ corresponds in this model a tower of Kaluza-Klein
states $X^{(n)}$ (KK-excitations),  whose masses are related to the size of the
compact extra dimension introduced and the mass of the SM particle via the
relation $m^2_{X^{(n)}} \approx m^2_{X^{(0)}}+n^2/R^2$. An important aspect of
the UED model is that it provides a viable candidate to the Cold Dark Matter.
This would be the lightest KK particle (LKP) which typically is the level 1
photon. Many aspects of the phenomenology of these KK excitations have been
discussed in the literature. For reviews see
ref.~\cite{Hooper:2007qk,Macesanu:2005jx,Rizzo:2010zf,Cheng:2010pt}. In
particular KK production has been considered both at the Cern large hadron
collider (LHC) and at the next linear collider (ILC). Direct searches of KK
level excitations at collider experiments give a current bound on the scale of
the extra-dimension of the order $R^{-1} \gtrsim 300$ GeV. See for example
ref.~\cite{PDBook}. At the Fermilab Tevatron it will be possible to test
compactification scales up to $R^{-1} \sim 500$ GeV at least within some
particular scenario~\cite{Macesanu:2002db,Macesanu:2002ew,Macesanu:2002hg}.

Lower bounds on the compactification radius arise also from analysis of
electro-weak precision measurements performed at the $Z$ pole (LEP II). An
important feature of these type of constraints is their  dependence on the Higgs
mass. A recent refined analysis~\cite{Gogoladze:2006br} taking into account
sub-leading contributions from the new physics  as well as two-loop corrections
to the standard model $\rho$ parameter finds that $R^{-1} \gtrsim 600$ GeV for a
light Higgs mass ($m_H= 115$ GeV) and a top quark  mass $m_t=173$ GeV at 90\%
confidence level (C.L.). Only assuming a larger value of the Higgs mass the
bound is considerably weakened down to $R^{-1} \gtrsim 300$ GeV for $m_H= 600$
GeV, thus keeping the model within the reach of the Tevatron run II. The finding
of this precision analysis are in qualitative agreement with previous
results~\cite{Appelquist:2002wb}, but are at variance with the conclusions of a
recent paper~\cite{Flacke:2005hb} where an analysis of LEP data including data
from above the $Z$ pole and two loop electro-weak corrections to the $\Delta
\rho$ parameter pointed to $R^{-1} \gtrsim 800$ (at 95\% C.L.).

 It has been shown in ref.~\cite{Haisch:2007vb}
that a refined analysis of $\bar{B}\to X_s\gamma$ including in addition to the leading order contribution
from the extra-dimensional KK states, the known next-to-next-to-leading order
correction in the Standard model (SM) gives a lower bound on the
compactification radius $R^{-1} \gtrsim 600$ GeV at $95\%$ confidence level (CL)
and independent of the Higgs mass.

We discuss here  the formation, production and possible detection  of bound 
states of Kaluza-Klein $n=1$ excitations  at $e^+e^-$ collisions.   
To estimate the bound state contribution to the threshold cross-section, an effect 
which can be as large as roughly a factor of three for  strongly interacting 
particles, we use the method of the Green function as opposed to previous
 works~\cite{Carone:2003ms} which use a Breit-Wigner approximation.  The 
 interactions responsible for the formation of level-1 KK bound states is  assumed to be
 an $\alpha_s$ driven Coulomb potential. This allows the  use of  analytic expressions
 for  the Green function of the Coulomb problem. This method has also been
recently used by the present authors in a study of \emph{sleptonium} bound
states within a slepton co-next to lightest supersymmetric particle (slepton
co-NLSP) scenario of gauge mediated symmetry breaking
(GMSB)~\cite{Fabiano:2005gt}.

\begin{table}
\begin{tabular}{cccccc}\hline
$R^{-1}$ (GeV)&$KK$ {mass} (GeV) & $\alpha_s(r_B^{-1})$&  $M_{\cal B}$ (GeV)& $E_{1S}$ (GeV) & $\Delta E(2P-1S)$ (GeV) \\
\hline
400 & 478.05 &0.131&952.57&   3.627  & 2.720 \\
600 & 717.08 &0.124&1429.30&  4.903  & 3.677 \\
800 & 956.11 &0.120&1906.11&  6.089  & 4.567 \\
1000& 1195.14 &0.116&2382.98& 7.214  & 5.411 \\\hline
\end{tabular}
\caption{\label{table1}Results of Coulombic model for the bound state of 
the level-1 iso-doublet $U_1$ quark. The strong coupling $\alpha_s $ is computed
at the scale $Q=r_B^{-1}$, where $r_B=3/(2m\alpha_s)$ is the Bohr's radius. For
each mass value  $m$ the scale $Q=r_B^{-1}$ depending itself on $\alpha_s$ must
be solved numerically from the equation $Q=(2/3)\, m \,\alpha_s(Q)$.}
\end{table} 

\section{ $\mathbf{u_1 \overline{u}_1}$ Bound State Formation and 
Production Cross-Section}
\label{sec:stateformation}

In this section we shall review the possible creation of a
bound state of the level-1  $KK$-excitation of the
$u$-quark, i.e. a bound state  $u_1 \bar{u}_1$. The
interaction among two Kaluza--Klein excitations are driven
by the QCD interaction, thus bearing no differences with
respect to the Standard Model; the strength of the
interaction is given by $\as$ computed at a suitable
scale~\cite{Fabiano:1993vx,Fabiano:1994cz,Fabiano:1997xh}.
We shall adopt the same formation criterion stated there,
namely that the formation occurs only if the level
splitting depending upon the relevant interaction existing
among constituent particles is larger than the natural
width of the would--be bound state. This translates into
the formation requirement \beq \Delta E_{2P-1S} \ge \Gamma
\label{eq:formation} \eeq where $\Delta E_{2P-1S} =
E_{2P}-E_{1S}$ and $\Gamma$ is the width of the would--be
bound state. The latter is twice the width of the single
$KK$ quark, $\Gamma = 2 \Gamma_{KK}$, as each $KK$ quark
could decay in a manner independent from the other.

In our model $V(\bm{x})$ is given by  a Coulombic potential
$ V(r) = - 4 \as/(3r)$
with $r=|\bm{x}|$, and
where $\as$ is the usual QCD coupling constant which
has been taken at a suitable scale as described
in~\cite{Fabiano:1993vx,Fabiano:1994cz}.
We are thus able to compute its energy levels given by the expression
\beq
\varepsilon_n = - \frac{4 }{9} \frac{m\as^2}{n^2}
\label{eq:ebound}
\eeq
and the separation of the first two energy levels is given by
$ \Delta E_{2P-1S} = m \as^2/3 $.
The scale at which $\as$ is evaluated is given by the inverse of
Bohr's radius $r_B=3/(2m\as)$, the average distance of the constituents of 
the bound state.
The mass of the $n$th bound state is
given by the expression $M_n = 2m+\varepsilon_n$
where $m$ is the mass of the constituent $u_1$ quark and $E_n$ is given
by~(\ref{eq:ebound}). The wavefunction at the origin, which will be needed
in order to compute decay widths, for this particular model is given
by the expression
$|\psi(0)|^2 = ( 2 m\as /3)^3/ \pi$. 
The results are given in Table~\ref{table1}.

In order to determine whether the bound state will be formed we shall apply the
criterion given in eq.~(\ref{eq:formation}). The $KK$-quark decay widths have
been already computed in~\cite{Carone:2003ms}, where it has been shown that
their values are at most of the order of 100~MeV, one order of magnitude less
than the energy splittings. In this scenario the eq.~(\ref{eq:formation})
requirement is always fulfilled, and the  bound state is formed for $KK$-quark
masses in this investigation range.

In order to describe the cross--section of a $KK$ bound
state in the threshold region we shall use the method of
the Green function. We briefly review here the essential
features of the mechanism, and refer the reader to the
literature for further details~\cite{Fabiano:2001cw}.
Let $\mathcal{G}_{1S}(\bm{x},\bm{y},E)$ be the Green function
of the Schr\"odinger equation which describes the
bound state by means of a suitable potential $V(\bm{x})$.

The complete expression for the $1S$ Green function of our problem
as a function of energy from threshold
is given with a slight change of notation by~\cite{Penin:1998mx}:
\begin{equation}
\mathcal{G}_{1S}(0,0,E+i\Gamma) = \frac{m}{4 \pi}
\left [-2 \lambda \left ( \frac{k}{2 \lambda}  +
\log \left( \frac{k}{\mu} \right ) + \psi(1-\nu) +2 \gamma-1  \right ) \right ]
\label{eq:green1s} \end{equation} where $ k = \sqrt{-m (E+i
\Gamma)}$, $ \lambda = 2 \as m/3$ and the wave number is $ \nu
=\lambda/k $; $E=\sqrt{s}-2m$. 
With the position $E \to E + i \Gamma$ we take into account the
finite width of the state.
The $\psi$ is the
logarithmic derivative of Euler's Gamma function
$\Gamma(x)$, $\gamma \simeq 0.57721$ is Euler's constant
and $\mu$ is an auxiliary parameter.

The final expression for the production cross--section of a $KK$ bound state  is
thus given by
\begin{equation}
\sigma(m,E,\Gamma,\as) = \frac{18 \pi}{m^2} \sigma_B\,\text{Im} \, \left[\mathcal{G}_{1S}\right]
\label{eq:sigmagreen}
\end{equation}
where $\sigma_B$ is the Born expression of the
cross--section~\cite{Burnell:2005hm} The process $e^+e^- \to U_1 \bar{U}_1$
proceeds through the annihilation into the standard model (level-0) gauge bosons
$\gamma $ and $Z$ but in principle one should also consider the contribution of
the  level-2 gauge bosons $\gamma_{(2)}$ and   $Z_{(2)}$.  Especially so in our
case of threshold production of the pair $u_1 \bar{u}_1$. Indeed in this case
$m\approx 1/R$, and $\sqrt{s} =2m +E \approx 2/R +E$ and since 
$m_{\gamma_2}\approx 2/R$ when producing at threshold the $u_1\bar{u}_1$ pair we
would be close  to the $\gamma_2$ and $Z_2$ resonances.  However the mass
spectrum is modified by the radiative corrections.  We have verified that over
the region of parameter space $ 300\,\,  \text{GeV }\leq R^{-1} \leq 1000\,\, 
\text{GeV}$ and  $2 \leq \Lambda R \leq 70$ the pair production threshold
$2m_{u_1}$ is always larger than $m_{\gamma_2}, m_{Z_2}$ and thus these
resonances should in principle be included in the calculation.  We have also
verified, cross checking our calculation with the output of a
CalcHEP~\cite{Pukhov:2004ca,Datta:2010us} session, that the numerical impact of
these diagrams is  completely negligible. Their contribution turns out to be
five orders of magnitude smaller than that of the SM gauge bosons $\gamma,Z$.
The analytic formula of the Born pair production cross section $e^+e^- \to
\gamma^*,Z^* \to U_1\bar{U}_1$ can be deduced for example from those  of heavy
quark ($t \bar{t}$) \cite{Panella:1993zk} taking into account the fact the the
level-1 KK quarks are \emph{vector-like} i.e. their coupling to the $Z$ is of
the $\gamma^\mu$ type and has no axial component.   Details of the calculation
and the explicit expression of the cross section are in~\cite{Fabiano:2008xk}.

In this work we shall concentrate on the continuum region of the cross--section,
namely $E>0$. The region below threshold, $E<0$, has been already discussed  in
ref.~\cite{Carone:2003ms}, using a Breit-Wigner description of both the 
positions and the widths of the peaks. In this respect the Green function
approach  is not expected to point to substantial differences relative to the
Breit--Wigner one. 
\begin{figure}[t!]
\includegraphics*[scale=0.795]{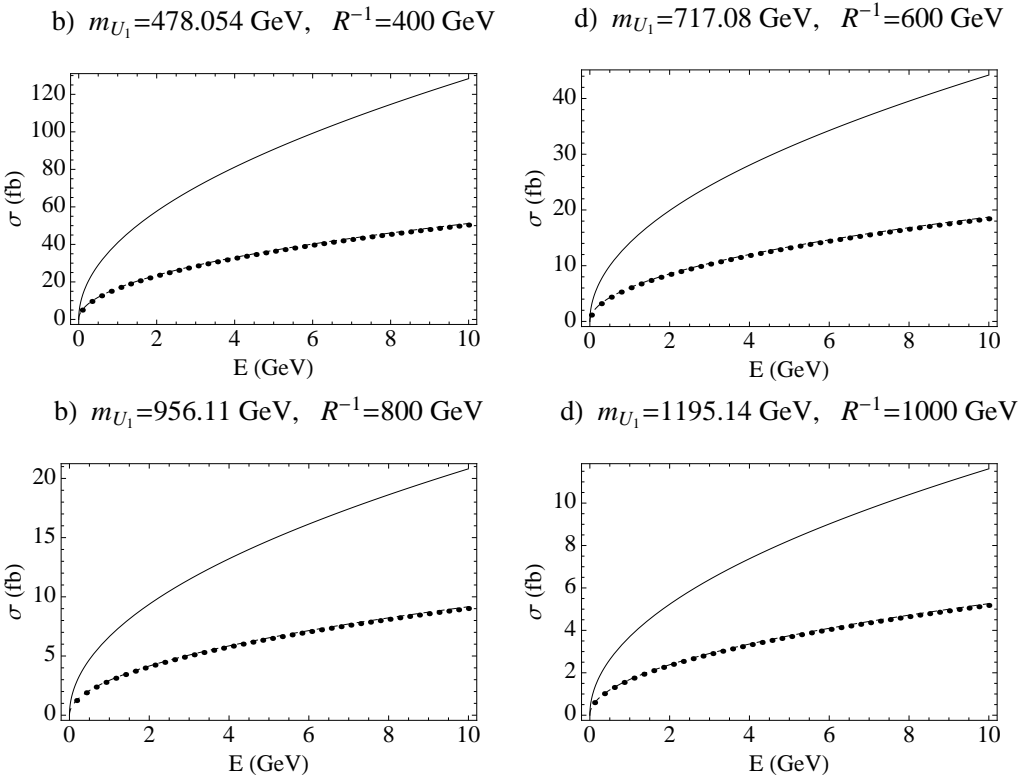}
\caption{\label{fig1}Production cross-sections  of level-1
KK doublet quark  bound states $U_1 \bar{U}_1$ as a function of the energy 
offset from threshold ($\sqrt{s} = 2m_{U_1} +E$), for values of the  scale 
of the extra-dimension $R^{-1}$ in the range $[400\div1000]$~GeV and a total 
width of $\Gamma=0.5$~GeV.  The continuous line is the Green Function result,
the dotted one is the Born approximation given by our analytical formula  
(see~\cite{Fabiano:2008xk}). The full circles show the complete agreement with 
 the Born cross section 
from  the CalcHEP~\cite{Datta:2010us} numerical session including also the 
annihilation diagrams of $\gamma_2$ and $Z_2$ whose contribution is however
completely negligible. 
The cut-off scale $\Lambda$, at which perturbative expansions break 
down, has been fixed so that $\Lambda R =20$.}
\end{figure}

In Fig.~\ref{fig1}  we  show the cross--section for selected values of the
scale of the extra dimension,  $R^{-1}=400-1000$~GeV. 
The results are
less sensitive to the other parameter ($\Lambda R$) which only enters through
the logarithmic factors in the radiative correction terms in the mass spectrum
of the model ~\cite{Fabiano:2008xk}. In fig.~\ref{fig1} we have fixed 
$\Lambda R=20$ and varied
$R^{-1}$ computing the corresponding values of the level-1 KK quark mass, and
assuming the energy of the collider being fixed at $\sqrt{s}=2m_{U_1}+ E$, $E$
being the energy offset from the threshold.  We have used a value of $\Gamma =
0.5$~GeV for illustrative purpose, compatible with the formation of bound state.
Different choices of $\Gamma$ by even two orders of magnitude smaller will not
make a visible difference on the figures.

\section{$\mathbf{u_1 \overline{u}_1}$ Decay Widths}
\label{sec:decays}
\begin{figure}[t!]
\begin{center}
\includegraphics*[scale=0.6]{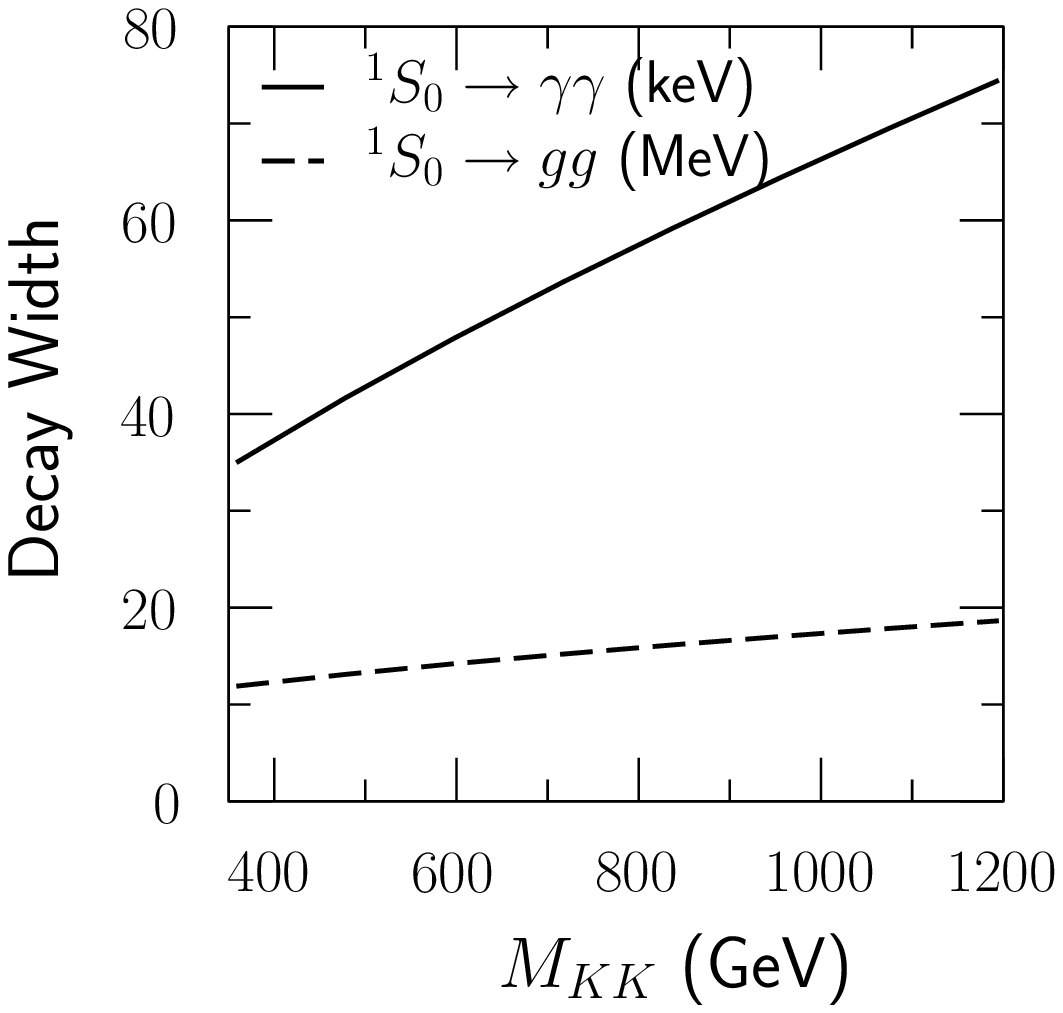}\hspace{0.5cm}\includegraphics*[scale=0.6]{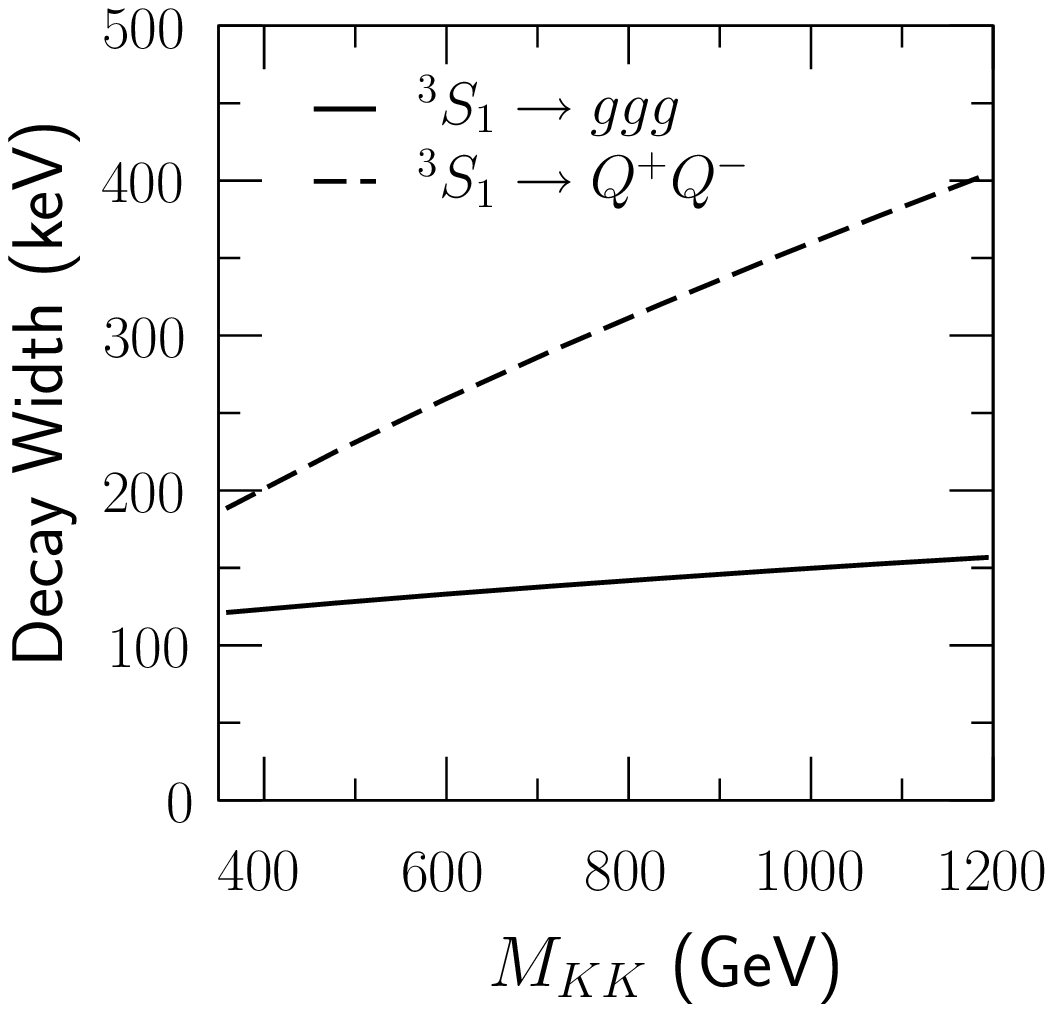}\end{center}
\caption{\label{fig2}\emph{\bf Left panel} Solid line: decay width (keV)
of the pseudoscalar $u_1 \bar{u}_1$ bound state to two
photons as a function of $KK$ excitation mass. \emph{Dashed
line}: decay width (MeV) of the pseudoscalar $u_1
\bar{u}_1$ bound state to two gluons as a function of the
$KK$ mass. \emph{\bf Right Panel}: decay widths of the $u_1 \bar{u}_1$
vector bound state to two charged particles and into three
gluons as a function of $KK$ mass. Here we have considered
all possible e.m. decay channels.
}
\end{figure}
The $KK$ bound states we discuss here are the pseudoscalar $^1S_0$ and
the vector one $^3S_1$. For the pseudoscalar state the decay channels are
into two photons or two gluons for which the following Born level expressions
hold (see for instance~\cite{Fabiano:1994cz}):
\beq
\Gamma_B(^1S_0 \to \gamma \gamma) =   q_i^4\alpha^2 \frac{48 \pi |\psi(0)|^2}{M^2}
\; \; \mathrm{and} \; \; \Gamma_B(^1S_0 \to gg) =   \as^2 \frac{32 \pi
|\psi(0)|^2}{3M^2} \; .
\label{eq:scalarto2photons}
\eeq
Here $q_i$ is the charge of the constituent quark of the bound state,
while $M$ and $|\psi(0)|^2$ are given by the ones from Coulombic potential
previously met.

The QCD radiative correction results~\cite{Kwong:1987ak}, which are the same
in the two decays, are shown in Fig.~\ref{fig2} (left panel).

For the vector case $^3S_1$ the relevant decay channels are the one in charged
pairs and the one into three gluons, for which one has
\beq
\Gamma_B(^3S_1 \to q_f^+q_f^-) =   q_i^2 q_f^2 \alpha^2 \frac{16 \pi |\psi(0)|^2}{M^2}
\; \; \mathrm{and} \; \; \Gamma_B(^3S_1 \to ggg) = \frac{(\pi^2-9)}{\pi}  \as^3
\frac{160 \pi |\psi(0)|^2}{81M^2} .
\label{eq:vectortoee}
\eeq
The charge of the final state charged particle is given by $q_f$.
The QCD radiative corrections~\cite{Kwong:1987ak} depend as usual on $\as$,
that has to
be computed at a scale of the order of $2m$.  The two decays of the vector state
are shown together in Fig.~\ref{fig2} (right panel). We observe that only the
pseudoscalar hadronic decay is in the MeV range and raises approximately
linearly with $KK$ mass. The $^1S_0$ photonic decay and $^3S_1$ decays are
smaller by almost two orders of magnitude for the considered $KK$ mass range.
For the pseudoscalar case the hadronic is the dominant decay by far, while in
the vector case the decay into charged particles, when taking into account all
possible processes as seen in Fig.~\ref{fig2} overtakes the hadronic decays.

Other electro-weak decay channels are negligible. Those are proportional to
$\alpha^2$, thus their ratio to gluonic decays is suppressed by
$(\alpha/\as)^2$, at least by two orders of magnitude.

For most scenarios depending upon the values of $\Lambda$ and
$R$~\cite{Carone:2003ms} single quark decay becomes the dominant decay channel
for the bound state.

Three--body decays are further suppressed with respect to previous formula
by another power in coupling constant and phase--space reduction, 
resorting again in the keV range of energies.

From~\cite{Carone:2003ms} one sees that in most cases single quark decays (SQD)
are by far the most important decay channels of the bound state, to the order
of hundreds of MeV, while as discussed above bound state decays are
essentially negligible.
Moreover a comparison of those SQD widths with the results of
Table~\ref{table1} through eq.~(\ref{eq:formation}) shows that
for the considered mass range of $KK$ there is formation of the
bound state.
\section{Detection}
\label{sec:detection}
As we have previously seen, for large $R^{-1}$ values ($R^{-1}> 300$~GeV) SQD is
the dominant decay channel for a $KK$ bound state, thus leading a to a dominant
signature consisting of two monochromatic quarks plus missing energy. 
Following~\cite{Fabiano:2001cw} we limit our analysis to the region above
threshold,  \emph{i.e.} $E>0$. The region below threshold, $E<0$,  is
characterized by peaks in the cross section for values of $E$ equal to binding
energies of the bound states. The width of those peaks are given by the decay
width of the bound state, which are at most of the order of the MeV for the SQD
and much less, of the order of the keV, for other annihilation decay modes, as
discussed in sect.~\ref{sec:decays}.

Because of ISR and beam energy spread, of the order of the GeV for a future
linear collider,
it is unclear whether it could be possible to resolve those peaks
of keV magnitude with this machine. The only potentially detectable peaks
should be the ones belonging to a SQD, provided one has a scenario with widths
of the order of the MeV.

The situation above threshold changes drastically with respect to the ``naive''
Breit--Wigner estimate, as is clearly shown in
Fig.~\ref{fig1}. A few GeV above threshold make
for a factor of 3 of increase compared to the Born cross--section, allowing
a clear distinction between the two cases. Assuming an annual integrated
luminosity of $L_0 = 100$ fb$^{-1}$ and a scale of the extra-dimension 
$R^{-1} =400$ GeV  one finds
around $1.2\times 10^4$ events per year of two quark decay for a center of mass
energy of 10 GeV above threshold (we adopt here the scenario for which the
branching ratio of SQD is essentially 1).
The number of events per year loses an order of magnitude at 
$R^{-1}=600$~GeV, that is about $4 \times 10^3$, as could be inferred
from Fig.~\ref{fig1}.

For our $\overline{u}_1 u_1$ bound state there are
two possible scenarios of decay
pattern~\cite{Cheng:2002ab}. The first one concerns the
iso-singlet $u_{1_R}$ for which the decay channel into
$W_1$ is forbidden while that into $Z_1$ is heavily
suppressed ${\cal B} (u_{1_R} \to Z_1 u_{0_R})\sim
\sin^2\theta_1 \approx 10^{-2} \div 10^{-3}$ and the
dominant channel is given by $u_{1_R} \to u_{0_R}
\gamma_1$, with ${\cal B}(u_{1_R} \to u_{0_R}
\gamma_1)\approx 0.98$ whose signature is a monochromatic
quark and missing energy of the $KK$ photon, the latter
being the LKP~\cite{Cheng:2002ab}.

For the iso-doublet $u_{1_L}$ the situation is more interesting, as more
channels are available~\cite{Cheng:2002ab}, notably $u_{1_L} \to d_{0_L} W_1$
with ${\cal B}(u_{1_L} \to d_{0_L} W_1) \approx 0.65$ and $u_{1_L} \to u_{0_L}
Z_1$, with ${\cal B}(u_{1_L} \to u_{0_L} Z_1) \approx 0.33$ while the branching
ratio into $\gamma_1$ is negligible ${\cal B}(u_{1_L} \to u_{0_L} \gamma_1)\sim
0.02$. The decay chain into $W_1$ can follow the scheme: $ u_{1_L} \to d_{0_L}
W_1 \to d_{0_L} \ell_0 \nu_{1_L} \to  d_{0_L} \ell_0 \nu_0 \gamma_1$ with
branching ratio given by:
\[
{\cal B}(u_{1_L}\to d_{0_L} \ell_0 \nu_0 \gamma_1)
\approx {\cal B}(u_{1_L} \to d_{0_L} W_1) {\cal B}(W_1 \to l_0 \nu_1)
\,{\cal B}( \nu_1 \to \nu_0\gamma_1) 
\approx 0.65\,\frac{1}{6} \, 1 \approx 10^{-1}
\]
and alternatively, the same final state could be reached by the scheme:
$ u_{1_L} \to d_{0_L} W_1 \to d_{0_L} \ell_1 \nu_{0_L} \to
d_{0_L} \ell_0 \nu_{0_L} \gamma_1 $. As compared to the iso-singlet case, the
result is a monochromatic quark, a \emph{lepton} and missing energy in both
cases.

The decay into the $Z_1$ channel is $ u_{1_L} \to u_{0_L} Z_1 \to u_{0_L} \ell_0
\ell_1 \to u_{0_L} \ell_0 \ell_0 \gamma_1$, resulting in a monochromatic quark,
two leptons and missing energy. The branching ratio of the above chain is:
\[
{\cal B}(u_{1_L}\to u_{0_L} \ell_0 \ell_0 \gamma_1)
\approx {\cal B}(u_{1_L} \to u_{0_L} Z_1)\, {\cal B}(Z_1
\to L_0 L_1) \,{\cal B}( L_1 \to
\ell_0\gamma_1)  \approx \frac{1}{3} \,\frac{1}{6}
\, 1 \approx 5\times 10^{-2}
\]
These leptonic decays of $u_1$ have much cleaner signatures
than the hadronic ones allowing, in principle, for a better
detection of the signal.

In all cases we emphasize that the observable signal of the
bound state production at the linear collider would be
similar to that of the Born pair production except for the
absolute value of the cross-section.   

The case of
an iso-singlet bound state (or Born pair production of
${u_1}_R$) would give rise to the signal
$
e^+e^- \to 2\, \textrm{jets} + E\!\!\!\! /
$
with cross section:
\begin{equation}
\sigma (e^+e^- \to 2\, \textrm{jets} + E\!\!\!\! /) \approx 
\sigma_{\cal B_{KK}}\times \left[{\cal B}({u_1}_R\to u_0
\gamma_1)\right]^2\, . 
\end{equation}
We note that the $\sigma_{\cal B_{KK}}$ for the iso-singlet $u_1$ has to be
computed ex-novo and cannot be read from the values of Fig.~\ref{fig1} since it
refers to the iso-doublet $U_1$. The singlet and doublet have, when including
radiative corrections, different masses and the corresponding pair production
threshold is therefore different. See ref.~\cite{Fabiano:2008xk}  for details.

At an $e^+e^-$ collider this signal has a standard model
background from $ZZ$ production with one $Z$ decaying to neutrinos and the other
decaying hadronically.   Thus for the  SM background  we have:
\begin{equation}
\sigma_{SM}(2\,\textrm{jets} +E\!\!\!\!/ ) = \sigma_{ZZ}\,
\text{fb}\times 0.7 \times 0.2 
\end{equation}
In the case of an iso-doublet bound state (or Born pair
production of ${U_1}$)  the $W_1$ decay chain gives  the
signal:
$ e^+e^- \to 2\, \textrm{jets} +2 \ell +  E\!\!\!\! / $
with cross section:
\begin{equation}
\sigma(e^+e^- \to 2 \textrm{j} +2 \ell +  E\!\!\!\! /)=
\sigma_{{\cal B}_{KK}} \left[{\cal B}(u_{1L}\to d_{0_L} \ell_0 \nu_0 \gamma_1 )\right]^2 \end{equation}
while  the $Z_1$ decay chain gives rise to the signature 
$ e^+e^- \to 2\, \textrm{jets} +4 \ell +  E\!\!\!\! / $
with cross sections:
\begin{equation}
\sigma(e^+e^- \to 2 \textrm{j} +4 \ell +  E\!\!\!\! /)=
\sigma_{{\cal B}_{KK}} \left[{\cal B}(u_{1L}\to u_{0_L}
\ell_0 \ell_0 \gamma_1 )\right]^2 \end{equation}
Triple gauge boson production, $WWZ, ZZZ$ at a high energy
linear collider has been studied in
refs.~\cite{Barger:1988sq,Barger:1988fd}. It has been found
that these processes receive a substantial enhancement in
the higgs mass range $200$ GeV $<m_H<600$ GeV particularly
the $ZZZ$ channel. As these processes provide a source of
standard model background for our signal we estimate them
both at a value of $m_h =120$ GeV and at a value of $m_h=200$ GeV 
for which the cross sections are
enhanced. Production of $WWZ$ can for instance give rise to
the signature of $2 \textrm{jets}+ 2 \ell + E\!\!\!\! / $
via leptonic decay of the W gauge bosons and hadronic decay
of the $Z$ boson, while the $ZZZ$ production can produce
$2\textrm{jets}+ 4 \ell + E\!\!\!\! / $ via hadronic decay
of one $Z$ while the others decay leptonically with one of them to a pair of
$\tau$ which subsequently decay to $\ell \nu\bar{\nu}$ ($\ell =e,\mu$).
Estimates of the resulting cross sections are found using the
CalcHEP~\cite{Pukhov:2004ca} and CompHEP~\cite{Boos:2004kh} software. We have
verified agreement with previous results  given in ref.~\cite{Barger:1988fd}.
We thus estimate within the standard model:
\begin{eqnarray}
\label{sigmabg2} \sigma_{\text{SM}}(2 \textrm{j}+ 2 \ell
+ E\!\!\!\! /) &\approx& \sigma_{WWZ}\times (0.1)^2\times
0.7 \cr \sigma_{\text{SM}}(2
\textrm{j}+ 4 \ell + E\!\!\!\! /) &\approx& \sigma_{ZZZ}\times(0.3)^2\times 0.7\times (0.17)^2 \end{eqnarray}

The $2 \textrm{jets}+ 2 \ell + E\!\!\!\! /$ channel could
be potentially contaminated also from $t\bar{t}$ pair
production cross section which at such high energies is
${\cal O}(300)$ fb~\cite{Weiglein:2004hn}. Assuming the top
quarks to decay with probability one to $Wb$ and then the
$W$ gauge boson decay via the leptonic mode (with ${\cal
B}(W\to \ell \nu_\ell) \approx 0.1$) would mimic the signal
with a cross section $\sigma_{SM}(2 \textrm{jets}+ 2 \ell +
E\!\!\!\! /) \approx 3 fb$. However in this case we expect
$b$-tagging of the hadronic jets. Assuming an efficiency in
$b$-tagging of $60\%$ we would get a contribution of $1.2$
fb  to the $2 \textrm{jets}+ 2 \ell + E\!\!\!\! /$
cross-section which has  to be added to that in
Eq.~\ref{sigmabg2}.  This has been done in the calculation
of the statistical significance of table~\ref{table2}.

We conclude providing an estimate of the statistical
significance
$
\label{statsig}
SS= {N_{\text{s}}}/{\sqrt{N_{\text{s}}+N_{\text{b}}}},
$
of the three signals discussed above as related to an
integrated luminosity of $L_0=100$ fb$^{-1}$ ($N_\text{s}$ is the number of 
signal events and $N_\text{b}$ is the number of background events). These
estimates are given in table~\ref{table2}.  Albeit quite
encouraging (especially so the $SS$ of the $2 \textrm{jets} + E\!\!\!\!
/ $) we should bear in mind that the actual observation of
these signals might be not be so easy from the experimental
point of view. Indeed it is quite likely that in a
framework of a quasi degenerate KK mass spectrum the jets
will be typically quite soft and therefore difficult to detect.
It is therefore customary to concentrate on the much
cleaner multilepton signatures~\cite{Cheng:2002ab,Battaglia:2005zf}. 
An analysis similar to the one given here, but with a
perspective on signals arising at the Compact Linear Collider (CLIC),  
regarding
the (Born) pair-production of level-1 KK-\emph{leptons} and level-1
KK-\emph{quarks} is given in~\cite{Battaglia:2005zf}.
\begin{table}
\begin{tabular}{ccccc}\hline
   $R^{-1}$ (GeV)& $m_{u_1}$ (GeV) &$2\, \textrm{jets} + E\!\!\!\!/$ &$2\, \textrm{jets}
   + 2\, \ell + E\!\!\!\!/ $&
   $2\, \textrm{jets} +4\, \ell+ E\!\!\!\!/$\cr
& (iso-singlet) & & $m_h=120$ (200) GeV & $m_h=120$ (200) GeV \cr\hline
400 & 469.0 &81.1 &8.2 (8.1) &5.6 (5.6) \cr
600 & 703.5 &44.4 &4.1 (4.1) &3.3 (3.3)\cr
800 & 938.0 &29.0 &2.4 (2.4) &2.3 (2.3)\cr
1000 & 1172.5 &20.8&1.6 (1.6) &1.7 (1.7)\cr\hline
 \end{tabular}
\caption{\label{table2}The statistical
significance $SS$ as defined in the text corresponding to 
the annual integrated luminosity $L_0=100$ fb$^{-1}$ for  the three channels 
discussed in the text as a function of  $R^{-1}$ and $\sqrt{s}=2m_{U_1} +E$,
assuming an energy offset of $E=10$ GeV from the threshold. 
In the second column we give the values of the $u_1$ iso-singlet
 level-1 KK quarks whose masses are different form those of the corresponding 
$U_1$ state from Fig.~\ref{fig1}.
For the two
multilepton channels $SS$ has been computed for two values of the Higgs mass
$m_h=120\, (200)\, \text{GeV}$  $(\Lambda R = 20)$.}
\end{table}

\section{Conclusions}
\label{sec:conclusions} 
We have considered the formation and decay of a bound state of level-1 quark
Kaluza-Klein excitation in UED and its consequent detection at a linear $e^+e^-$
collider. Since $m_{KK}$ should be larger than at least 300~GeV we have used a
model with a Coulombic Being a bound state we have used the Green function
technique for the evaluation of its formation cross--section in the threshold
region, which is more appropriate than the standard Breit--Wigner picture as it
takes into account the binding energy and the peaks of the higher level
excitations that coalesce towards the threshold point. The net effect is a
dramatic increase of the cross--section in the continuum region right of the
threshold. This multiplicative factor is roughly 2.6 for $R^{-1}=400 $ GeV and
drops down to 2.2 at $R^{-1}=1000$ GeV. The Green function cross-section would
allow more than $\approx 10^4$ events per year even at $R^{-1}=400$ GeV
($m_{U_1}\approx 478$~GeV) for a suitable integrated luminosity of the $e^+e^-$
linear collider ($L_0 =100$ fb$^{-1}$). The number of events at $R^{-1}=1000$
GeV ($m_{U_1}\approx 1200$~GeV) would still be $\approx 10^3$ at the same
integrated luminosity.

The large difference among the two descriptions of the cross--section should
also possibly help in the determination of the correct model for such a heavy
bound state outside the SM.

Our analysis of the backgrounds to the final states signals, though very
simplified, indicates that the multi-lepton channels have a good statistical
significance ($SS \gtrsim 2$) at least up to $R^{-1} =600 \sim 700$ GeV,  which
certainly warrants further detailed and dedicated studies of these channels and
their backgrounds.  The potentially larger  (by one order of magnitude) 
statistical significance of the $2j +E\!\!\!\! / $\ channel must be taken
however  with great caution because this signal may be difficult to observe as
it is  characterized by soft jets within the relatively degenerate mass spectrum
of the extra-dimensional model. Further detailed studies are also needed for
this channel. 
\begin{acknowledgments}
N.~F. was supported by the {\scshape Fondazione Cassa di Risparmio di Spoleto}.
\end{acknowledgments}


\begin{thebibliography}{0}

\bibitem{Kaluza:1921tu}
\BY{T.~Kaluza} \IN{Sitzungsber, Preuss. Akad. Wiss. Berlin (Math. Phys. )}{966}{1921}{}

\bibitem{Klein:1926tv}
O.~Klein, Z. Phys. \textbf{37}, 895 (1926)

\bibitem{Antoniadis:1990ew}
I.~Antoniadis, Phys. Lett. \textbf{B246}, 377 (1990)

\bibitem{Antoniadis:1998ig}
I.~Antoniadis, N.~Arkani-Hamed, S.~Dimopoulos, G.R. Dvali, Phys. Lett.  \textbf{B436}, 257 (1998), \texttt{hep-ph/9804398}

\bibitem{ArkaniHamed:1998rs}
N.~Arkani-Hamed, S.~Dimopoulos, G.R. Dvali, Phys. Lett. \textbf{B429}, 263
  (1998), \texttt{hep-ph/9803315}

\bibitem{Randall:1999ee}
L.~Randall, R.~Sundrum, Phys. Rev. Lett. \textbf{83}, 3370 (1999),
  \texttt{hep-ph/9905221}

\bibitem{Randall:1999vf}
L.~Randall, R.~Sundrum, Phys. Rev. Lett. \textbf{83}, 4690 (1999),
  \texttt{hep-th/9906064}

\bibitem{Appelquist:2000nn}
T.~Appelquist, H.C. Cheng, B.A. Dobrescu, Phys. Rev. \textbf{D64}, 035002
  (2001), \texttt{hep-ph/0012100}

\bibitem{Hooper:2007qk}
D.~Hooper, S.~Profumo, Phys. Rept. \textbf{453}, 29 (2007),
  \texttt{hep-ph/0701197}

\bibitem{Macesanu:2005jx}
C.~Macesanu, Int. J. Mod. Phys. \textbf{A21}, 2259 (2006),
  \texttt{hep-ph/0510418}


\bibitem{Rizzo:2010zf}
  T.~G.~Rizzo,
  arXiv:1003.1698 [hep-ph].

\bibitem{Cheng:2010pt}
  H.~C.~Cheng,
  arXiv:1003.1162 [hep-ph].

\bibitem{PDBook}
W.M. {Yao}, C.~{Amsler}, D.~{Asner}, R.~{Barnett}, J.~{Beringer}, P.~{Burchat},
  C.~{Carone}, C.~{Caso}, O.~{Dahl}, G.~{D'Ambrosio} et~al., {Journal of
  Physics G} \textbf{33}, 1+ (2006), \texttt{http://pdg.lbl.gov}

\bibitem{Macesanu:2002db}
C.~Macesanu, C.D. McMullen, S.~Nandi, Phys. Rev. \textbf{D66}, 015009 (2002),
  \texttt{hep-ph/0201300}

\bibitem{Macesanu:2002ew}
C.~Macesanu, C.D. McMullen, S.~Nandi, Phys. Lett. \textbf{B546}, 253 (2002),
  \texttt{hep-ph/0207269}

\bibitem{Macesanu:2002hg}
C.~Macesanu, C.D. McMullen, S.~Nandi (2002), \texttt{hep-ph/0211419}

\bibitem{Gogoladze:2006br}
I.~Gogoladze, C.~Macesanu, Phys. Rev. \textbf{D74}, 093012 (2006),
  \texttt{hep-ph/0605207}

\bibitem{Appelquist:2002wb}
T.~Appelquist, H.U. Yee, Phys. Rev. \textbf{D67}, 055002 (2003),
  \texttt{hep-ph/0211023}

\bibitem{Flacke:2005hb}
T.~Flacke, D.~Hooper, J.~March-Russell, Phys. Rev. \textbf{D73}, 095002 (2006),
  \texttt{hep-ph/0509352}

\bibitem{Haisch:2007vb}
U.~Haisch, A.~Weiler, Phys. Rev. \textbf{D76}, 034014 (2007),
  \texttt{hep-ph/0703064}

\bibitem{Carone:2003ms}
C.D. Carone, J.M. Conroy, M.~Sher, I.~Turan, Phys. Rev. \textbf{D69}, 074018
  (2004), \texttt{hep-ph/0312055}

\bibitem{Fabiano:2005gt}
N.~Fabiano, O.~Panella, Phys. Rev. \textbf{D72}, 015005 (2005),
  \texttt{hep-ph/0503231}

\bibitem{Papavassiliou:2001be}
J.~Papavassiliou, A.~Santamaria, Phys. Rev. \textbf{D63}, 125014 (2001),
  \texttt{hep-ph/0102019}

\bibitem{Cheng:2002ab}
  H.~C.~Cheng, K.~T.~Matchev and M.~Schmaltz,
  Phys.\ Rev.\  D {\bf 66}, 056006 (2002)
  [arXiv:hep-ph/0205314].


\bibitem{Fabiano:1993vx}
N.~Fabiano, A.~Grau, G.~Pancheri, Phys. Rev. \textbf{D50}, 3173 (1994)

\bibitem{Fabiano:1994cz}
N.~Fabiano, G.~Pancheri, A.~Grau, Nuovo Cim. \textbf{A107}, 2789 (1994)

\bibitem{Fabiano:1997xh}
N.~Fabiano, Eur. Phys. J. \textbf{C2}, 345 (1998), \texttt{hep-ph/9704261}

\bibitem{Fabiano:2001cw}
N.~Fabiano, Eur. Phys. J. \textbf{C19}, 547 (2001), \texttt{hep-ph/0103006}

\bibitem{Datta:2010us}
  A.~Datta, K.~Kong and K.~T.~Matchev,
  arXiv:1002.4624 [hep-ph].

\bibitem{Penin:1998mx}
A.A. Penin, A.A. Pivovarov, Phys. Atom. Nucl. \textbf{64}, 275 (2001),
  \texttt{hep-ph/9904278}

\bibitem{Burnell:2005hm}
F.~Burnell, G.D. Kribs, Phys. Rev. \textbf{D73}, 015001 (2006),
  \texttt{hep-ph/0509118}

\bibitem{Pukhov:2004ca}
  A.~Pukhov,
  arXiv:hep-ph/0412191.

\bibitem{Panella:1993zk}
  O.~Panella, G.~Pancheri and Y.~N.~Srivastava,
  Phys.\ Lett.\  B {\bf 318} (1993) 241.

\bibitem{Fabiano:2008xk}
  N.~Fabiano and O.~Panella,
  Phys.\ Rev.\  D {\bf 81} (2010) 115001
  [arXiv:0804.3917 [hep-ph]].


\bibitem{Kwong:1987ak}
W.~Kwong, P.B. Mackenzie, R.~Rosenfeld, J.L. Rosner, Phys. Rev. \textbf{D37},
  3210 (1988)


\bibitem{Barger:1988sq}
V.~D.~Barger and T.~Han,
Phys.\ Lett.\  B {\bf 212}, 117 (1988).

\bibitem{Barger:1988fd}
  V.~D.~Barger, T.~Han and R.~J.~N.~Phillips,
  Phys.\ Rev.\  D {\bf 39}, 146 (1989).
\bibitem{Boos:2004kh}
  E.~Boos {\it et al.}  [CompHEP Collaboration],
  Nucl.\ Instrum.\ Meth.\  A {\bf 534}, 250 (2004)
  [arXiv:hep-ph/0403113].
\bibitem{Weiglein:2004hn}
  G.~Weiglein {\it et al.}  [LHC/LC Study Group],
  Phys.\ Rept.\  {\bf 426}, 47 (2006)
  [arXiv:hep-ph/0410364].

\bibitem{Battaglia:2005zf}
  M.~Battaglia, A.K.~Datta, A.~De Roeck, K.~Kong and K.~T.~Matchev,
  JHEP {\bf 0507}, 033 (2005)
  [arXiv:hep-ph/0502041].

\end{thebibliography}
\end{document}